\documentstyle[12pt]{article}
\def\simge{\hspace*{0.2em}\raisebox{0.5ex}{$>$}
     \hspace{-0.8em}\raisebox{-0.3em}{$\sim$}\hspace*{0.2em}}
\def\simle{\hspace*{0.2em}\raisebox{0.5ex}{$<$}
     \hspace{-0.8em}\raisebox{-0.3em}{$\sim$}\hspace*{0.2em}}

\def\sst#1{{\scriptscriptstyle #1}}

\def\sst#1{{\scriptscriptstyle #1}}

\def\lamchi{{\Lambda_\chi}}

\def\qw0{{Q_\sst{W}^0}}

\def\gaf{{g_\sst{A}^f}}

\def\qpv{{Q_\sst{W}}}

\def\mz{{M_\sst{Z}}}
\def\mzs{{M^2_\sst{Z}}}

\def\mpi{{m_\pi}}

\def\lamchi{{\Lambda_\chi}}

\def\lamtv{{\Lambda_\sst{TVPC}}}
\def\lamtvs{{\Lambda_\sst{TVPC}^2}}
\def\lamtvc{{\Lambda_\sst{TVPC}^3}}

\def\osffp{{{\cal O}_7^{ff'}}}
\def\osgg{{{\cal O}_7^{\gamma g}}}
\def\osgz{{{\cal O}_7^{\gamma Z}}}

\begin{document}

\begin{titlepage}

\begin{center}

{\large{\bf Searching for T-Violating, P-Conserving New Physics with Neutrons}}

\vspace{1.2cm}

M. J. Ramsey-Musolf$^{a}$

\vspace{0.8cm}

$^a$ Department of Physics, University of Connecticut,
Storrs, CT 06269 USA\\

\end{center}

\vspace{1.0cm}
\begin{abstract}
The observance of parity conserving time reversal violation in light quark systems
could signal the presence of physics beyond the Standard Model. I discuss the
implications of low-energy time reversal tests for the existence of such T-violating,
P-conserving (TVPC) interactions. I argue that searches for permanent electric dipole
moments (EDM's) and direct TVPC searches provide complementary information on
P-conserving T-violation. EDM searches yield constraints only 
under the assumption that parity symmetry is restored at the scale $\lamtv$
associated with new TVPC physics. If parity remains broken at short distances, direct
searches yield the least ambiguous bounds. In the latter case, improving the
experimental precision of direct TVPC searches in neutron $\beta$-decay and polarized
epithermal neutron transmission at the Spallation Neutron Source could yield tighter
bounds.

\vskip 0.5 true cm

\end{abstract}

\vspace{2cm}
\vfill
\end{titlepage}

\pagenumbering{arabic}

\section{Introduction}

The recent results for $K^0$-${\bar K}^0$ oscillations reported by CPLEAR have
provided the first direct evidence for the violation of time reversal
invariance\cite{Ang98}. Indirect evidence has existed since the discovery of
CP-violation in the neutral kaon system. In accordance with the CPT theorem of
quantum field theory, the existence of CP-violation implies the existence of
T-violation. An interesting question is whether there exist other direct signatures
of T-violation (TV) outside the neutral kaon system. Searches for TV in light quark
systems have constituted an ongoing topic in nuclear physics for the past quarter
century or more. So far, experimental studies of detailed balance in nuclear
reactions,  correlations in $\gamma$- and $\beta$-decays of nuclei (also the neutron
in the latter case), neutron and atomic electric dipole moments (EDM's), and
correlations in the scattering of epithermal neutrons from heavy nuclei have yielded
null results\cite{Boe95}. In some cases, these experimental limits place important
constraints on scenarios for physics beyond the Standard Model, making TV searches in
nuclear physics of interest to particle physicists as well.

The prospect of a new source of polarized neutrons at the Spallation Neutron Source
(SNS) suggests that neutron physics could be an ideal arena in which to carry out
future searches for TV. In this talk, I want to consider the use of neutrons to
search for TV \lq\lq new physics" which is parity-conserving (PC). As shown be
Herczeg {\em et al.}, one cannot write down a renormalizable gauge theory in which
TVPC interactions among quarks arise at tree-level from boson exchange \cite{Her}.
Thus, the existence of such interactions would signal the presence of some exotic
physics beyond the Standard Model. To arrive at such interactions, one would have to
forego the requirement that all interactions must be renormalizable or imagine some
kind of nonperturbative effects generating effective, nonrenormalizable interactions.
Indeed, an interesting challenge for theorists is to arrive at some plausible
mechanism for generating such TVPC interactions from some fundamental, underlying
theory. 

In this talk, I will focus on a more pedestrian issue, namely, the
phenomenological constraints on the existence of TVPC interactions. Several years
ago, it was suggested in Ref. \cite{Con92} that one could obtain constraints on
TVPC interactions from EDM's, even though existence of the latter require both
TV and parity-violation (PV). Moreover, it was argued in Ref. \cite{Con92} that 
the EDM constraints far exceed those obtained from direct TV searches. If
correct, this argument implies that one might as well forego the latter and
concentrate solely on the EDM. I hope to convince you, however, that the
arguments of Ref. \cite{Con92} are not airtight. The apply under only one of
several possible scenarios for the breakdown of parity symmetry. There exist
equally plausible scenarios under which EDM's are of limited value at best in
constraining new TVPC interactions.

My arguments depend on the use of ideas in effective field theory (EFT), which is
the only framework I know of for treating non-renormalizable interactions in a
systematic way when one does not have access to a fundamental short-distance
theory. In a nutshell, the argument may be stated as follows. Let $\lamtv$ denote
the mass scale below which an EFT using nonrenormalizable TVPC interactions is
sensible. One may then expand the EDM $d$ of a neutron, electron, or neutral atom
in powers of $1/\lamtv$:
\begin{equation}
\label{eq:dexpand}
{d\over e}=
\beta_5 C_5{1\over\lamtv}+\beta_6 C_6{M\over\lamtvs}+
\beta_7 C_7{M^2\over\lamtvc}+\cdots\ \ \ ,
\end{equation}
where the $C_n$ denote the set of {\em a priori} unknown coefficients of dimension
$n$ nonrenormalizable operators in the effective Lagrangian, the $\beta_n$ are
calculable quantities arising from  loops or hadronic matrix elements,
and $M<<\lamtv $ is a mass scale associated with any dynamical degree of freedom
in the EFT. The first contributions from new TVPC interactions appear in the
$C_7$. 

\noindent Now consider two scenarios: 

\medskip

\noindent {\bf Scenario (A)} Parity symmetry is restored at some scale
$\mu\simle\lamtv$. In this case, it turns out that all of the coefficients 
$C_5$ and $C_6$ must vanish. The first contributions to the EDM arise from
loop effects involving the TVPC $C_7$ operators. Since $M/\lamtv <1$, these
contributions presumably dominate the series. Consequently, one could use
EDM limits to place constraints on $C_7/\lamtvc$. This is the scenario implicitly
considered in Ref. \cite{Con92}. As I show below, the limits obtained from
EDM's in this case vastly exceed those obtainable from direct searches.

\medskip

\noindent {\bf Scenario (B)} Parity symmetry remains broken at $\mu\simge\lamtv$. In
this case, the $C_5$ and $C_6$ do not, in general, vanish. The EDM expansion contains
important contributions from lower dimension TVPV operators. Without independent
information about the $C_{5,6}$, one cannot use the EDM as a direct window on the
TVPC $C_7$ terms. The latter may be more or less suppressed relative to the 
lower dimension contributions depending on the size of $M/\lamtv$. Since we have
no {\em a priori} information on $M/\lamtv$ -- it may be $1/2$ or $1/20,\ldots$ --
we can say very little about the importance of the TVPC contributions. We might
assume that $M/\lamtv<<1$ so that the first term in Eq. (\ref{eq:dexpand})
presumably dominates. In this case, the low-energy effects of TVPC interactions
would have to be negligible. In the more general situation, however, one would
have to use direct TVPC searches to constrain the new TVPC interactions if this
scenario holds. Here, then, is the window of relevance for new TVPC searches with
neutrons. 

In the remainder of this talk, I flesh out these arguments in more detail,
summarizing my work in Ref. \cite{MRM99}. First, I review the ideas of EFT and
make reference to its more familiar use in chiral perturbation theory. 
I subsequently apply these ideas to the analysis of TVPC interactions and EDM's,
leading to the conclusions summarized above. Finally, I make some comments about
the phenomenology of time-reversal tests.

\section{Effective Field Theory}

Effective field theory is a powerful tool when our knowledge of dynamics above
some scale is limited. In the case of hadronic interactions, we known how to
handle strong meson-baryon interactions at energies well below the scale of
chiral symmetry breaking, $\lamchi\sim 1$ GeV. Our ability to calculate the
dynamics at shorter distances, where one must consider quark/gluon
substructure, is embryonic at best. Chiral perturbation theory (CPT) allows us
to circumvent our ignorance by parameterizing short distance QCD in terms of
an infinite tower of non-renormalizable interactions with {\em a priori} unknown
coefficients. This EFT is tractable so long as the energy of any process one is
interested in, and the masses of particles involved, is sufficiently below 
$\lamchi$. In this case, one may expand any observable in powers of $p/\lamchi$,
where $p$ is the appropriate momentum scale. One uses data to determine the
unknown coefficients, or low energy constants (LEC's). So long as there exists
enough available data to determine the relevant LEC's, one may then use chiral
symmetry to make predictions for new observables not previously considered.
Corrections to the symmetry relations arise from loop effects. The latter
generally scale as $m/\lamchi$ relative to the tree-level terms, where $m$ is
the mass of one of the particles in a loop. 

As an example, consider the octet baryon magnetic moments. In the nonrelativistic
version of CPT (heavy baryon CPT), magnetic moments arise at tree-level in a
Lagrangian containing dimension 5 operators:
\begin{equation}
\label{eq:mm1}
{\cal L}_5 = {b\over 2\lamchi}\epsilon_{\mu\nu\alpha\beta}{\bar B} v^\alpha
S^\beta B F^{\mu\nu}\ \ \ ,
\end{equation}
where $F^{\mu\nu}$ is the electromagnetic field strength tensor, $B$ is the field
for a baryon having velocity $v^\alpha$ and spin $S^\beta$. The coefficient $b$,
which is {\em a priori} unknown, contains all the information we lack on the short
distance QCD dynamics of magnetic moments. If the mass scale $\lamchi$ were
chosen correctly, then we should find this coefficient to have \lq\lq natural
size", $b\sim 1$. At this level of the theory, $B$'s magnetic moment is simply
related to $b$ as
\begin{equation}
\label{eq:mm2}
\mu_B=\left({M_B\over\lamchi}\right)b\ \ \ ,
\end{equation}
where $M_B$ is the baryon mass. If $b\sim 1$ and $\mu_B\sim 1$ as is the case for
octet baryon magnetic moments, then one should find $\lamchi\sim M_B\sim 1$ GeV.
From purely pionic interactions, we learn that $\lamchi=4\pi F_\pi\sim 1$ GeV, where
$F_\pi$ is the pion decay constant. At this level, then, the EFT appears to be
self consistent.

Corrections to the relation in Eq. (\ref{eq:mm2}) arise from loops involving pions
and baryons. These loops diverge quadratically. Na\"\i vely, they receive
contributions from arbitrarily high momentum scales. However, the assumption in
writing down this EFT is that we only know how to calculate the dynamics for
scales below $\lamchi$. In short, we rely on the assumption of a {\em separation
of scales}: physics for $p<\lamchi$ is calculable in the theory while physics
at scales $p\geq\lamchi$ is uncalculable but parameterized by the LEC's like $b$.
In order to ensure that our loop calculation does not blur this separation of
scales, we must regulate the loop in such a way that it receives contributions
only from intermediate states having $p<\lamchi$. An appropriate regulator for this
purpose is dimensional regularization (DR). After computing the regulated loops, we
obtain a new expression for $\mu_B$:
\begin{equation}
\label{eq:mm3}
\mu_B=\left({M_B\over\lamchi}\right)b(\mu)+\left({M_B\over\lamchi}\right)
\left({\mpi\over\lamchi}\right)g_A^2 {\tilde b(\mu)}\ \ \ .
\end{equation}
Here, $g_A$ the appropriate $\pi$-baryon coupling (or combination of couplings)
and $\tilde b$ is a loop-dependent number. After loops renormalize the tree-level
theory, all the LEC's in general contain an implicit dependence on the scale 
$\mu$ at which the integrals are defined. In some cases, the loop factors
$\tilde b$ contain no explicit $\mu$-dependence; in others, it is at most
a logarithmic dependence. 

The important feature from our standpoint is that 
the first term in the RHS of Eq. (\ref{eq:mm3}) contains the short-distance
physics we cannot calculate while the second term contains calculable long range
contributions from scales below $\lamchi$. This scale separation appears in the
guise of the $\mpi/\lamchi$ suppression factor in the second term. A factor like
this must appear since the operator of interest is dimension five (the magnetic moment
interaction) while the loop is quadratically divergent. Some power of mass must
appear in the numerator of the loop result. The scale separation of EFT implies
it can only be associated with a particle lighter than $\lamchi$. This \lq\lq
power counting" implies that loop effects are generally suppressed relative to
tree-level contributions, since $\mpi/\lamchi<1$. Moreover, we are allowed to
truncate the chiral expansion of at any order we choose, since higher-order
contributions -- arising either from additional loops and/or higher-dimension
operators -- will be smaller by additional powers of $\mpi/\lamchi$. 

It is worth emphasizing here that one would not use the magnetic moment expansion
of Eq. (\ref{eq:mm3}) to determine the value of the $g_A$ from magnetic moment
data. Since no symmetry rules out the presence of the leading term in the series,
it is the most important unknown in the expansion. One must use independent
observables, such as $\pi N$ scattering, to determine $g_A$. Only if some physics
consideration told us $b$ had to vanish would the magnetic moments give one a
direct handle on $g_A$. 

\section{Effective Field Theory for TVPC Physics}

Following the ideas of the previous section, we may expand the Lagrangian
containing new TV interactions in powers of $1/\lamtv$:
\begin{equation}
\label{eq:lnew}
{\cal L}_\sst{NEW} = {\cal L}_4 +{1\over\lamtv}{\cal L}_5+{1\over\lamtvs}{\cal L}_6
+{1\over\lamtvc}{\cal L}_7+\cdots \ \ \ .
\end{equation} 
Here, the subscripts denote the dimension of operators appearing in ${\cal L}_n$.
These operators are built out of fields representing elementary particles having
masses $<\lamtv$. The term ${\cal L}_5$ contains TVPV operators, including the
familiar EDM operator for an elementary fermion:
\begin{equation}
\label{eq:edmep}
{\cal O}_5 = -\frac{i}{2} C_5^f {\bar\psi}\sigma_{\mu\nu}\gamma_5\psi
\ F^{\mu\nu}\ \ \ .
\end{equation}
The operators in ${\cal L}_6$ are also TVPV. New TVPC interactions appear in
${\cal L}_7$. Among those of interest to us are the operators
\begin{eqnarray}
\label{eq:osffp}
{\cal O}_7^{ff'}&=&
C_7^{ff'} {\bar\psi}_f {\buildrel \leftrightarrow \over D_\mu} \gamma_5 \psi_f
	   {\bar\psi}_{f'}\gamma^\mu\gamma_5 \psi_{f'}\\
\label{eq:osgg}
{\cal O}_7^{\gamma g} &=& C_7^{\gamma g} {\bar\psi}\sigma_{\mu\nu}\lambda^a\psi
    F^{\mu\lambda} G_\lambda^{a\ \nu}\\
\label{eq:oszg}
{\cal O}_7^{\gamma Z} &=& C_7^{\gamma Z} {\bar \psi} \sigma_{\mu\nu}\psi\
F^{\mu\lambda}
     Z_\lambda^\nu\ \ \ .
\end{eqnarray}
Here, $G_\lambda^{a\ \nu}$ is the field strength tensor for a gluon with SU(3)
index \lq\lq a" and $Z_\lambda^\nu$ is the $Z^0$ boson field strength tensor.
For simplicity, I will consider only neutral gauge bosons here. The operator
$\osffp$ was first considered in Ref. \cite{Con92}, while $\osgg$ was introduced
in Ref. \cite{Eng}. The $Z-\gamma$ TVPC operator $\osgz$ appeared first in Ref.
\cite{MRM99}. In addition to these operators, there also exist dimension seven
TVPV operators in ${\cal L}_7$. 

The operators appearing in Eq. (\ref{eq:lnew}) will contribute to various TVPC and/or
TVPV observables, such as the neutron EDM. The presently uncalculable short distance
($p\geq\lamtv$) contributions to these observables live in the operator coefficients,
$C_n$. The calculable long distance ($p<\lamtv$) effects arise from loops or
many-body matrix elements containing the various ${\cal O}_n$. Now we may consider the
implications of the EFT expansion of Eq. (\ref{eq:lnew}) under the two scenarios
outlined in the introduction.

Consider first scenario (B), where parity violation remains in effect for
$p\simge\lamtv$. In this case, the coefficients of the TVPV operators do not, in
general, vanish. There exists no reason why any new  short distance TVPC physics
cannot conspire with short distance PV interactions to generate non-vanishing
coefficients of these operators. This statement is analogous to saying that the
coefficient of the leading term in the magnetic moment expansion, $b$, does not
vanish. Nothing about the uncalculable short distance QCD physics of magnetic moments
forbids the existence of this term. In the case of scenario (B), then, the leading
contribution to the neutron EDM arises from the ${\cal O}_5$. In the relativistic
quark model, one has
\cite{MRM99}
\begin{equation}
\label{eq:quarkmodel}
d_n={1\over\lamtv}\int\ d^3x (u^2+\frac{1}{3}\ell^2)\ \left[\frac{4}{3} C_5^d -
\frac{1}{3} C_5^u\right]\ \ \ ,
\end{equation}
where $u$ and $\ell$ denote the upper and lower components, respectively, of the
ground state quark wavefunction. In the language of Eq. (\ref{eq:dexpand}), $\beta_5$
is just the integral in Eq. (\ref{eq:quarkmodel}) and $C_5$ is the linear combination
of quark EDM's, $C_5^{u,d}$, multiplying the integral (the factor of $e$ is buried in
these constants). Numerically, one has $\beta\sim 0.88$. 

Higher-order contributions to $d_n$ arise from the TVPV operators in ${\cal L}_6$,
${\cal L}_7$, {\em etc.} as well as from the TVPC operators in ${\cal L}_7$ and
beyond. The latter contribute via loop graphs which contain a PV interaction. For
example, $\osgz$ contributes via one-loop diagrams in which the virtual $Z^0$ has
a PV interaction with the fermion involved. In the leading log approximation, the
result is\cite{MRM99}
\begin{equation}
\label{eq:oneloop}
C_5^{f,\ \sst{LOOP}}\sim e C_7^{\gamma Z} \left({\mz\over\lamtv}\right)^2\left({1\over
s_W c_W}\right)
   \gaf\left({1\over 16\pi^2}\right) \ln{\mzs\over\mu^2} \ \ \ ,
\end{equation}
where $\mu$ is the DR subtraction scale and $\gaf$ is the axial vector
$Zff$ coupling\footnote{Note that the result in Eq. (\ref{eq:oneloop}) is a factor
of 6 larger than the formula given in Ref.
\cite{MRM99}.}. The most important feature of this result is the appearance of the
$(\mz/\lamtv)^2$ factor, as required by the power counting of EFT. Since $\mz$ must
be smaller than $\lamtv$ in order for an EFT expansion of
${\cal L}_\sst{NEW}$ to make sense, the contribution of $\osgz$ to the neutron EDM is
suppressed relative to the leading term. The size of this suppression depends on the
degree of separation between $\mz$ and $\lamtv$. The additional loop factors further
suppress this contribution:
\begin{equation}
{1\over 16\pi^2}\ln{\mzs\over\mu^2}\sim {1\over 10}
\end{equation}
when one takes $\mu\sim 1$ GeV as is appropriate for the neutron EDM. Thus, even if
$\mz$ and $\lamtv$ were comparable, the contribution of the TVPC operators to $d_n$
would be smaller than the tree-level contributions of the lower-dimension TVPV
operators. One simply cannot use the EDM to constrain the new TVPC interactions in
this case. 

If scenario (A) holds, however, all of the TVPV coefficients in Eq.
(\ref{eq:lnew}) must vanish at tree-level. Since there exists no short distance PV,
the existence of these operators is forbidden by parity symmetry. In this case,
the leading contribution to $d_n$ arises from the loops containing the $d=7$ TVPC
operators. Consequently, experimental EDM limits place strong constraints on the
ratios $C_7/\lamtvc$. These limits imply that -- if scenario (A) holds --
low-energy TVPC effects should be incredibly small. Specifically, let $\alpha_T$
denote the  ratio of typical nuclear TVPC nuclear matrix elements to those of the
residual strong interaction. On dimensional grounds, this ratio should scale as
\begin{equation}
\label{eq:alphatscale}
\alpha_T\sim C_7(p/\lamtv)^3\ \ \ ,
\end{equation} 
where $p$ is a typical momentum for a low-energy TVPC process. Taking $p= 1$
GeV$/c$ and experimental bounds on the neutron EDM, one would expect $\alpha_T
\simle 10^{-13}$. 

I emphasize that the arguments of Ref. \cite{Con92} apply only to scenario (A)
and not to scenario (B). Even in the case of its application to scenario (A),
however, the analysis of Ref. \cite{Con92} does not respect the separation of scales
underlying the use of EFT. Instead of using DR to regulate their loop integrals, the
authors of Ref. \cite{Con92} employed a cut-off regulator, where the cut-off scale
was taken to be $\lamtv$! Consequently, the loops in that calculation are dominated by
intermediate states having momenta $p\sim\lamtv$, and the result does not contain the
$(\mz/\lamtv)^2$ factor implied by EFT power counting. As shown in Ref. \cite{MRM99},
using a cut-off regulator in this way has disatrous consequences: TVPC operators of
arbitrarily high dimension contribute to the EDM with the same weight as the leading
$d=7$ TVPC operators. One has no way, then, to truncate the series and --
correspondingly -- no way to derive an experimental bound on anything other than an
infinite series of comparably weighted nonrenormalizable operators. In contrast, the
use of DR -- which preserves the power counting of EFT -- affords one with a
systematic truncation scheme. Only by having the latter can one hope to derive
meaningful limits on new TVPC interactions from EDM bounds under scenario (A).

\section{Mass Scales and Phenomenology} 

It is often instructive to interpret limits on new physics in terms of a mass scale
associated with some symmetry breakdown. For example, the Boulder group \cite{Wie}
has recently determined the weak charge, $\qpv$, of the cesium atom to about 0.35\%
experimental error. When combined with the up-dated estimate of the atomic theory
uncertainty\cite{Wie}, this measurement of $\qpv$ is sensitive to the existence of
right handed neutral gauge bosons as heavy as $\sim 1$ TeV\cite{MRM99b}. One would
then conclude that the scale of left-right symmetry breaking should not be
considerably below one TeV.

One might similarly ask what time-reversal tests teach us about the mass scale
associated with the breakdown of time-reversal invariance. Since we do not yet have a
detailed theory of how T might be broken unaccompanied by P violation, the best we
can do is use dimensional and \lq\lq naturalness" arguments to derive approximate
bounds on $\lamtv$. The problem is our ignorance of the dynamics which could generate
the $C_n$ in Eq. (\ref{eq:lnew}). Nevertheless, we might parameterize our ignorance
in the following way. Let
\begin{equation}
\label{eq:nat1}
C_7^{ff'} = 4\pi\kappa^2\ \ \ ,
\end{equation}
where $\kappa$ characterizes the coupling strength of new TVPC interactions. Since
$\osgz$ contains a $\gamma$ and $Z^0$, its coefficient should contain additional
factors of $e$ and $g$ as compared to $C_7^{ff'}$. This choice would be the \lq\lq
natural" one. Thus we would have
\begin{equation}
\label{eq:nat2}
C_7^{\gamma Z}\sim egC_7^{ff'}={4\pi\alpha\over\sin\theta_\sst{W}} C_7^{ff'}.
\end{equation}

Now consider the experimental limits on the electric dipole of the electron. The most
recent measurement by the Berkeley group gives\cite{Com94}
\begin{equation}
|d_e| < 4\times 10^{-27} \ \ e\ \mbox{cm}\ \ \ .
\end{equation}
Under scenario (A), this result constrains the TVPC interactions. The most stringent
bounds apply to $\osgz$, which contribute at one-loop order. Using Eqs.
(\ref{eq:nat1}, \ref{eq:nat2}), we obtain
\begin{equation}
\label{eq:bound1}
\lamtv\simge 270 \kappa^{2/3} \ \ \mbox{TeV}
\end{equation}
from the $d_e$ bounds\footnote{This bound is $6^{1/3}$ stronger than quoted in Ref.
\cite{MRM99} due to the factor of six in the loop expression mentioned previously.}.  
If the new TVPC physics is of a strong interaction origin, then $\kappa\sim 1$ and
the scale of T breakdown would be at least three times as heavy as the weak scale.
Note that parity invariance would have to be borken somewhere below this scale in
order for this limit to be valid.

The limits on $\lamtv$ under scenario (B) are considerably weaker. These limits
arise from the direct TVPC searches mentioned in the Introduction. One may perform
a simple-minded estimate using Eq. (\ref{eq:alphatscale}) and the direct search
bounds: $\alpha_T\simle 10^{-3}$. For $p\sim 1$ GeV$/c$, one would obtain the bound
\begin{equation}
\label{eq:bound2}
\lamtv\simge 20\kappa^{2/3}\ \ \ \mbox{GeV}\ \ \ .
\end{equation}
A detailed analysis of various TVPC observables suggests that the limit may actually
be weaker than Eq. (\ref{eq:bound2}) \cite{GCM00}. Since TVPC observables depend on 
$1/\lamtvc$, it would take an improvement of at least 12 orders of magnitude in
experimental precision to push the scenario (B) limits to the level of the scenario
(A) limits. A more realistic goal may be to determine whether there exists any window
for new TVPC physics to exist below the weak scale\cite{comment}. Reaching the weak
scale would require improvement of three or more orders of magnitude in experimental
precision. To this end, the direct TVPC searches with neutrons at the SNS may be one
avenue to pursue. Whether such improvements are feasible in measurements of the
$\beta$-decay $D$ coefficient\cite{Lis00} or the five-fold correlation in epithermal
neutron-nucleus transmission\cite{Gou99} is a challenge for experimentalists. In this
respect, another interesting possibility is the search for charge symmetry-breaking
TVPC terms in neutron-proton elastic scattering or a five-fold correlation in
proton-deuteron transmission\cite{vanO}.

\section{Conclusions}

Tests of fundamental symmetries could form a significant component of the physics
program at the SNS. Studies of PV in neutron $\beta$-decay and in neutron spin
rotation could provide new insight into the nature of the
semileptonic weak interaction and low-energy strong interaction, respectively. 
Improvements in neutron EDM limits will further constrain gauge theories containing
both PV and TV. In this talk, I hope I have convinced you that there exists room for
improvement in searches for new TVPC physics as well. The relationship between the
scale at which new TVPC physics arises and the scale at which parity symmetry is
restored is not clear at present. What one can say is that if parity symmetry is
restored by the time one reaches $\lamtv$, EDM's imply that $\lamtv$ is well above
the weak scale. On the other hand, if PV remains in effect at $\lamtv$, direct
TVPC searches give us the least ambiguous window on this type of new physics.


\begin{thebibliography}{99}
\bibitem{comment} If $\lamtv$ is below the weak scale, the loop calculation yielding
Eq. (\ref{eq:oneloop}) would not apply, since it contained a dynamical $Z^0$ boson.
\end{thebibliography}
\end{document}